\documentclass[aip,cha,preprint,amsmath,amsthm,amssymb,showpacs,floatfix,graphicx]{revtex4-1}
\usepackage{graphicx}
\usepackage{dcolumn}
\usepackage{bm}
\newcommand{\be}{\begin{equation}}
\newcommand{\ee}{\end{equation}}

\newtheorem{theorem}{Theorem}

\newtheorem{conj}[theorem]{Conjecture}

\begin{document}
\title{ 
Stability of Fixed Points and Chaos in Fractional Systems
}

\author{Mark Edelman}

\affiliation{Department of Physics, Stern College at Yeshiva University, 245
  Lexington Ave, New York, NY 10016, USA 
\\ Courant Institute of
Mathematical Sciences, New York University, 251 Mercer St., New York, NY
10012, USA\\
Department of Mathematics, BCC, CUNY, 2155 University Avenue, 
Bronx, New York 10453
}

\date{\today}

\begin{abstract}
In this paper we propose a method to define the range of stability of
fixed points for a variety of discrete fractional systems of the order
$0 < \alpha <2$. The method is tested on various forms of fractional 
generalizations of the standard and logistic maps. 
Based on our analysis we make a conjecture that chaos
is impossible in the corresponding continuous fractional systems.
\end{abstract}

\maketitle

{\bf Many natural (biological, physical, etc.) and social systems posess  
power-law memory  and can be described by the fractional 
differential/difference equations. Nonlinearity is an important property
of these systems. Behavior of such systems can be very different from
the behavior of the correcponding systems with no memory. Previous reserch
on the issues of the first bifurcations and the stability of 
fractional systems mostly adddressed the question of sufficient
conditions. In this paper we propose the equations that allow calculations of
the coordinates of the asymptotically stable period two sinks and the
values of nonlinearity and memory parameters defining the first
bifurcation form the stable fixed points to the $T=2$ sinks. 
}


\section{Introduction}
\label{sec:1}

It is generally understood that socioeconomic and biological systems are
systems with memory. Specific analysis showing that the memory in
financial and socioeconomic systems obeys the power law can be found in 
papers \cite{MachadoFinance,Machado2015,TarasovEconomic} and sources cited 
in these papres. Power-law in human memory was investigated in
\cite{Kahana,Rubin,Wixted1,Wixted2,Adaptation1,Donkin}:
the accuracy on memory tasks decays as a power law  
$\sim t^{-\beta}$, with $0<\beta<1$  and, with respect to human learning, 
it is shown in \cite{Anderson} that the reduction in reaction times that 
comes with practice is a power function of the number of training trials.
Power-law adaptation has been used 
to describe the dynamics of biological systems in papers
\cite{Adaptation1,Adaptation3,Adaptation4,Adaptation2,Adaptation5,Adaptation6}.
 
The impotence and origin of the memory in biological systems can be 
related to the presence of memory at the level of individual cells:  
it has been shown recently that processing of external stimuli by 
individual neurons 
can be described by fractional differentiation \cite{Neuron3,Neuron4,Neuron5}. 
The orders of fractional derivatives $\alpha$ derived for different types of 
neurons fall within the interval [0,1], which implies power-law memory 
$\sim t^{\beta}$ with power $\beta = 1-\alpha$, $\beta \in [-1,0]$. 
For neocortical pyramidal neurons the order of the fractional derivative 
is quite small: $\alpha \approx 0.15$.

Viscoelastic properties of the human organ 
tissues are best described by fractional differential equations with
time fractional derivatives, which implies the power-law memory
(see, e.g., references in \cite{Chaos2015}). 
In most of the biological systems with the power-law behavior 
the power $\beta$ is between -1 and 1 ($0 < \alpha <2$).

Among the fundamental scientific problems driving interest and research in 
fractional dynamics are the origin of memory and 
a possibility of memory being present in the very basic equations of
Physics. 
Could it be that the fundamental laws describing fields and particles are
not memoryless and are governed by 
fractional differential/difference equations? 

\begin{figure*}[]
\begin{center}
\includegraphics[width=0.6\textwidth]{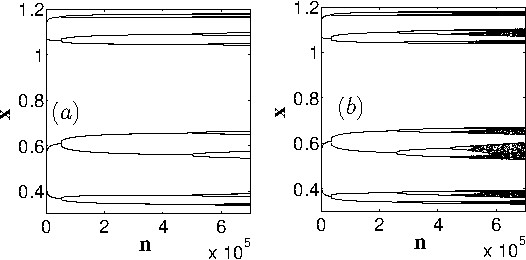}
\vspace{-0.25cm}
\caption{Two examples of cascade of bifurcations type trajectories in the
Caputo logistic $\alpha$-family of maps (Eq.(\ref{FrCMapx}) with $h=1$ and  
$G_K(x)=x-Kx(1-x)$ ) with $\alpha=0.1$ and $x_0=0.001$: 
(a) for the nonlinearity parameter $K=22.37$ the last bifurcation from the period  
$T=8$ to the period $T=16$ occurs after approximately $5 \times 10^5$ iterations; 
(b) when  $K=22.423$ the trajectory becomes chaotic after approximately 
$5 \times 10^5$ iterations.
}
\end{center}
\label{fig:1}
\end{figure*}

Because most of the social, biological, and physical systems are
nonlinear, it is important to look for the fundamental differences in 
the behavior of nonlinear systems with and without memory. 
Let's list some of the differences.
 \begin{itemize}
\item
{
Trajectories in continuous fractional systems of orders less than two
may intersect (see, e.g., Fig.~2 form \cite{Chaos2015}) and chaotic 
attractors may overlap (see, e.g., Fig.~4~f from \cite{ME3}).
}
\item
{As a result of the previous statement, the Poincar$\acute{e}$-Bendixson 
Theorem does not apply to fractional systems and even in continuous
systems of the order $\alpha<2$ non-existence of chaos is only a
conjecture (see \cite{Chaos2015,Deshpande}). 
}
\item
{ 
Periodic sinks may exist only in asymptotic sense and asymptotically 
attracting points may not belong to their own basins of attraction 
(see \cite{ME3,ME2,ME4}). A trajectory starting from an asymptotically 
attracting 
point jumps out of this point and may end up in a different
asymptotically attracting point.  
}
\item
{A way in which a trajectory is approaching an attracting point depends
on its origin. Trajectories originating from the basin of attraction may 
converge faster (as $x_n \sim n^{-1-\alpha}$ for the fractional 
Riemann-Liuoville standard map, see Fig.~1 from \cite{ME3}) than
trajectories 
originating from the chaotic sea (as $x_n \sim n^{-\alpha}$).
}
\item
{Cascade of bifurcations type trajectories (CBTT) exist only in fractional 
systems. The periodicity of such trajectories is changing with time: they
may start converging to the period $2^n$ sink, but then bifurcate and
start converging to the period $2^{n+1}$ sink and so on. CBTT may end its 
evolutions converging to the period $2^{n+m}$ sink (Fig. ~\ref{fig:1}(a))   or in chaos (Fig. ~\ref{fig:1}(b)) \cite{ME2,Chaos}.
}
\item
{Continuous and discrete fractional systems may not have periodic 
solutions except fixed points (see, e.g., 
\cite{PerD1,PerD2,PerC1,PerC2,PerC3,PerC4,PerC5}. 
Instead they may have asymptotically periodic solutions.
}
\item
{Fractional extensions of the volume preserving systems are not volume 
preserving. If the order of a fractional system is less than 
the order of
the corresponding integer system, then 
behavior of the system is similar to the behavior of the corresponding
integer system with dissipation \cite{ZSE}. 
Correspondingly, the types of attractors which may 
exist in fractional systems include sinks, limiting cycles, and chaotic 
attractors \cite{ME4,ME5,DNC,AttrC1,AttrC2}}
\end{itemize}
A particular problem related to the differentiation between fractional systems 
and integer ones, the first bifurcation on CBTT, and related problems of 
stability of fixed points in discrete fractional systems and transition to 
chaos in continuous fractional systems are considered in this paper.

Stability of fractional systems was investigated in numerous papers based
on various methods (Lyapunov's direct and indirect methods, Lyapunov
function, Routh-Hurwitz criterion, ...). Here we'll list only some of
the research papers, reviews, and books on the topic. Paper
\cite{Matignon} is the most cited article on stability of linear
fractional differential equations. In application to stability of 
nonlinear fractional differential equations, we'll mention papers 
\cite{PRL,Ahmed,ElSaka,StL,StA,StLW,StLB}. Some of the results on
stability of discrete fractional systems can be found in papers 
\cite{StanislavskyMaps,StChen,StJarad,StMohan,StDis1,StB}. 
The reviews on the topic include papers \cite{Rev2009,Rev2011,Rev2013} 
and books \cite{Petras,Zhou}. Almost all results obtained in the cited
papers define sufficient conditions of stability and don't allow to
calculate the ranges of nonlinearity parameters and orders of derivatives
for which fixed points are stable.   

In this paper we derive the algebraic equations to calculate asymptotically period two sinks of discrete fractional systems, which define the conditions of their appearance, 
and conjecture that these equations define the values of 
nonlinearity parameters and orders of derivatives
for which fixed points become unstable. This conjecture is 
numerically verified for the fractional standard and logistic maps.
This paper is a continuation of the research on general properties of 
fractional systems based on the properties of fractional maps 
\cite{Chaos2015,ME3,ME2,Chaos,ME4,ME5,DNC,StanislavskyMaps,Chaos2014,ME6,ME9,ME10,T2009a,T2009b,T2,T1,Fall}.
In Sec.~\ref{sec:2} we review the most common forms of fractional maps. In 
Sec.~\ref{sec:3} we derive the equations defining the ranges of nonlinearity 
parameters and orders of derivatives for which fixed points are stable. 
Sec.~\ref{sec:4} presents the summary of our results.

\section{Fractional/fractional difference maps}
\label{sec:2}

In this section some essential definitions and theorems are presented.

\subsection{Fractional integrals and derivatives}
\label{sec:2.1}

In this paper we will use the definition of fractional integral introdused 
by Liouville, which is a generalization of the Cauchy formula for the 
n-fold integral
{\setlength\arraycolsep{0.5pt}
\begin{equation}
 _aI^{p}_t x(t) 
=\frac{1}{\Gamma(p)} \int^{t}_a 
\frac{x(\tau) d \tau}{(t-\tau)^{1-p}}~,
\label{RLI}
\end{equation}
}
where $p$ is a real number, $\Gamma()$ is the gamma function and we'll assume $a=0$.

The left-sided Riemann-Liouville fractional derivative $_0D^{\alpha}_tx(t)$ 
is defined for $t>0$ \cite{KST,Podlubny,SKM} as 
{\setlength\arraycolsep{0.5pt}
\begin{eqnarray} 
&&_0D^{\alpha}_t x(t)=D^n_t \ _0I^{n-\alpha}_t x(t)  \nonumber \\
&&=\frac{1}{\Gamma(n-\alpha)} \frac{d^n}{dt^n} \int^{t}_0 
\frac{x(\tau) d \tau}{(t-\tau)^{\alpha-n+1}}~,
\label{RL}
\end{eqnarray} 
}
where $n-1 \le \alpha < n$, $n \in \mathbb{Z}$,  
$D^n_t=d^n/dt^n$.        

In the definition of the left-sided Caputo derivative, 
the order of integration and
differentiation in 
Eq. (\ref{RL}) is switched
\cite{KST}  
{\setlength\arraycolsep{0.5pt}
\begin{eqnarray} 
&&_0^CD^{\alpha}_t x(t)=_0I^{n-\alpha}_t \ D^n_t x(t) \nonumber \\
&&=\frac{1}{\Gamma(n-\alpha)}  \int^{t}_0 
\frac{ D^n_{\tau}x(\tau) d \tau}{(t-\tau)^{\alpha-n+1}},  \quad (n-1 <\alpha \le n).
\label{Cap}
\end{eqnarray} 
}

\subsection{Fractional sums and differences}
\label{sec:2.2}

We'll also use the proposed by Miller and Ross generalization of the forward 
sum/difference operator  \cite{MR} 
 \begin{equation} 
\Delta x(t)=x(t+1)-x(t) 
\label{FDiff}
\end{equation}
(see below) and call it simply the fractional sum/difference
operator. Nabla fractional difference, which is the generalization of 
the backward sum/difference operator $\nabla x(t)=x(t)-x(t-1)$ 
\cite{GZ1988} is not considered in this paper.

The fractional 
sum ($\alpha>0$)/difference ($\alpha<0$) operator deined in   \cite{MR} 
\begin{equation}
_a\Delta^{-\alpha}_{t}f(t)=\frac{1}{\Gamma(\alpha)} \sum^{t-\alpha}_{s=a}(t-s-1)^{(\alpha-1)} f(s)
\label{MRDef}
\end{equation}
is a fractional generalization of the $n$-fold summation formula 
\cite{ME9,GZ1988}
{\setlength\arraycolsep{0.5pt}   
\begin{eqnarray} 
&&_a\Delta^{-n}_{t}f(t)=\frac{1}{(n-1)!} \sum^{t-n}_{s=a}(t-s-1)^{(n-1)}
f(s) \nonumber \\
&&=\sum^{t-n}_{s^0=a} \sum^{s^0}_{s^1=a}...
\sum^{s^{n-2}}_{s^{n-1}=a}f(s^{n-1}),
\label{MRInt}
\end{eqnarray}
}
where $n \in \mathbb{N}$. In Eq.~(\ref{MRDef}) $f$ is defined on  $\mathbb{N}_a$ and $_a\Delta^{-\alpha}_t$ on  
$\mathbb{N}_{a+\alpha}$, where   $\mathbb{N}_t=\{t,t+1, t+2, ...\}$.
The falling factorial $t^{(\alpha)}$ is defined as
\begin{equation}
t^{(\alpha)} =\frac{\Gamma(t+1)}{\Gamma(t+1-\alpha)}, \ \ t\ne -1, -2, -3,
...
\label{FrFac}
\end{equation}
and is asymptotically a power function:
\begin{equation}
\lim_{t \rightarrow
  \infty}\frac{\Gamma(t+1)}{\Gamma(t+1-\alpha)t^{\alpha}}=1,  
\ \ \ \alpha \in  \mathbb{R}.
\label{GammaLimit}
\end{equation}

For $\alpha >0$ and   $m-1<\alpha \le m$ the fractional (left) 
Riemann-Liouville difference operator is defined (see \cite{Atici1,Atici2}) as
{\setlength\arraycolsep{0.5pt}   
\begin{eqnarray} 
&&_a\Delta^{\alpha}_t x(t) =  \Delta^{m} _a\Delta^{-(m-\alpha)}_{t}x(t) 
\nonumber \\
&&=\frac{1}{\Gamma(m-\alpha)} \Delta^m \sum^{t-(m-\alpha)}_{s=a}(t-s-1)^{(m-\alpha-1)} x(s)
\label{FDRL}
\end{eqnarray}
}
and the fractional (left) Caputo-like difference operator (see \cite{Anastas}) as
{\setlength\arraycolsep{0.5pt}   
\begin{eqnarray} 
&&_a^C\Delta^{\alpha}_t x(t) =  _a\Delta^{-(m-\alpha)}_{t}\Delta^{m} x(t)
\nonumber \\
&&=\frac{1}{\Gamma(m-\alpha)} \sum^{t-(m-\alpha)}_{s=a}(t-s-1)^{(m-\alpha-1)} 
\Delta^m x(s).
\label{FDC}
\end{eqnarray}
}
Due to the fact that $_a\Delta^{\lambda}_t$ in the limit 
$\lambda \rightarrow 0$ approaches the identity operator (see \cite{ME9,MR}), the 
definition Eq.~(\ref{FDC}) can be extended to all real $\alpha \ge 0$
with $_a^C\Delta^{m}_t x(t) = \Delta^m x(t)$ for $m \in \mathbb{N}_0$.

Fractional h-difference operators, which are generalizations of the 
fractional difference operators, were introduced in \cite{hdif1,hdif3}. 
The h-sum operator is defined as
\begin{equation}
(_a\Delta^{-\alpha}_{h}f)(t)=\frac{h}{\Gamma(\alpha)} \sum^{\frac{t}{h}-\alpha}_{s=\frac{a}{h}}(t-(s+1)h)^{(\alpha-1)}_h f(sh),
\label{hSum}
\end{equation}
where $\alpha \ge 0$, $(_a\Delta^{0}_{h}f)(t)=f(t)$, $f$ is defined on
$(h\mathbb{N})_a$, and $_a\Delta^{-\alpha}_h$ on  
$(h\mathbb{N})_{a+\alpha h}$. $(h\mathbb{N})_t=\{t,t+h, t+2h, ...\}$.
The $h$-factorial $t^{(\alpha)}_h$ is defined as
\begin{equation}
t^{(\alpha)}_h
=h^{\alpha}\frac{\Gamma(\frac{t}{h}+1)}{\Gamma(\frac{t}{h}+1-\alpha)}=
h^{\alpha}\Bigl(\frac{t}{h}\Bigr)^{(\alpha)}, 
\label{hFrFac}
\end{equation}
where  $t/h \ne -1, -2, -3, ...$.
With $m=\lceil \alpha \rceil$   the Riemann-Liouville (left) h-difference is defined as
{\setlength\arraycolsep{0.5pt}
\begin{eqnarray}
&&(_a\Delta^{\alpha}_h x)(t) =  (\Delta^{m}_h
(_a\Delta^{-(m-\alpha)}_{h}x))(t) =\frac{h}{\Gamma(m-\alpha)} \nonumber \\  
&& \times \Delta^m_h \sum^{\frac{t}{h}-(m-\alpha)}_{s=\frac{a}{h}}(t-(s+1)h)^{(m-\alpha-1)}_h 
x(sh)
\label{hFDRL}
\end{eqnarray}
}
and  the Caputo (left) h-difference is defined as
{\setlength\arraycolsep{0.5pt}
\begin{eqnarray}
&&(_a\Delta^{\alpha}_{h,*} x)(t) =  
(_a\Delta^{-(m-\alpha)}_h (\Delta^{m}_{h}x))(t)=\frac{h}{\Gamma(m-\alpha)}
 \nonumber \\  
&&\times \sum^{\frac{t}{h}-(m-\alpha)}_{s=\frac{a}{h}}(t-(s+1)h)^{(m-\alpha-1)}_h 
(\Delta^m_hx)(sh),
\label{hFDC}
\end{eqnarray}
}
where $(\Delta^{m}_{h}x))(t)$ is the $m$th power of the forward $h$-difference operator 
\begin{equation} 
(\Delta_{h}x)(t)=\frac{x(t+h)-x(t)}{h}.
\label{FHD}
\end{equation} 

As it has been noted in \cite{hdif1,hdif3}, due to the convergence of
solutions of fractional Riemann-Liouville h-difference equations 
when $h \rightarrow 0$ to 
solutions of the corresponding differential 
equations, they can be used
to solve fractional Riemann-Liouville differential equations
numerically. 


\subsection{Fractional maps}
\label{sec:2.3}

Maps with power-law memory can be introduced directly as a particular form of maps with memory (see papers
\cite{Chaos2015,StanislavskyMaps} which  contain references and 
discussions on the topic).
The most general form of the convolution-type map with power-law memory 
introduced in \cite{Chaos2015} can be written as
{\setlength\arraycolsep{0.5pt}
\begin{eqnarray}
&&x_{n}=\sum^{\lceil \alpha \rceil - 1}_{k=1}\frac{c_k}{\Gamma(\alpha-k+1)}
(nh)^{\alpha-k}  \nonumber \\  
&&+\frac{h^{\alpha}}{\Gamma({\alpha})}\sum^{n-1}_{k=0}(n-k)^{\alpha-1} G_K(x_k),
\label{EqDir}
\end{eqnarray}
where $\alpha \ge 0$, $K$ is a parameter, and $h$ is a constant 
time step between the
time instants $t_n=a+nh$ and $t_{n+1}$. For a physical interpretation 
of this formula we consider a system which state is defined by the 
variable $x(t)$ and evolution by the continuous function $G_K(x)$. 
The value of the state variable at the time $t_n$, $x_n=x(t_n)$, 
is a weighted total of the functions $G_K(x_k)$ from the values of 
this variable at the past time 
instants $t_k=a+kh$, $0 \le k<n$, $t_k=kh$. 
The weights are the times between the time instants $t_n$ and $t_k$ 
to the fractional power $\alpha-1$.
Eq.~ (\ref{EqDir}) in the limit $h \rightarrow 0+$ 
yields the Volterra integral equation of the second kind 
{\setlength\arraycolsep{0.5pt}
\begin{eqnarray}
&&x(t)= \sum^{\lceil \alpha
  \rceil-1}_{k=1}\frac{c_k}{\Gamma(\alpha-k+1)}(t-a)^{\alpha-k}
\nonumber \\  
&&+\frac{1}{\Gamma (\alpha)}
\int^{t}_{a}\frac{G_K(\tau,x(\tau))d\tau}{(t-\tau)^{1-\alpha}}. \  \ \ (t>a), 
\label{VoltRealNN}
\end{eqnarray}
This equation is equivalent to the fractional differential equation with
the Riemann-Liouville or Gr$\ddot{u}$nvald-Letnikov fractional 
derivative \cite{Chaos2015,KBT1,KBT2}
\begin{equation}
_a^{RL/GL}D^{\alpha}_tx(t)=G_k(t,x(t)), \  \ 0 <\alpha
\label{KMRL}
\end{equation}
with the initial conditions 
\begin{equation}
(_a^{RL/GL}D^{\alpha-k}_tx) (a+)=c_k, \  \  \ k=1,2,...,\lceil \alpha \rceil.
\label{IC3}
\end{equation}
For $\alpha \not\in \mathbb{N}$ we assume $c_{\lceil \alpha \rceil}=0$, 
which corresponds to a finite value of $x(a)$.

The same map, Eq.~(\ref{EqDir}), called the universal map,	 represents the solution of the fractional generalization of the differential equation of a periodically (with the period $h$) kicked system (see \cite{DNC,Chaos,ME5,T2009a,T2009b,T2,T1} for the fractional universal maps and \cite{ZasBook} in regular dynamics).

To derive the equations of the fractional universal map, which we'll call
the universal $\alpha$-family of maps ($\alpha$-FM) for $\alpha \ge 0$, 
we start with the differential equation 
\begin{equation}
\frac{d^{\alpha}x}{dt^{\alpha}}+G_K(x(t- \Delta h)) \sum^{\infty}_{k=-\infty} \delta \Bigl(\frac{t}{h}-(k+\varepsilon)
\Bigr)=0,   
\label{UM1D2Ddif}
\end{equation}
where $\varepsilon > \Delta > 0$,  $\alpha \in \mathbb{R}$, $\alpha>0$,
and consider 
it as $\varepsilon  \rightarrow 0$. The initial conditions
should correspond to the type of the fractional derivative
used in Eq.~(\ref{UM1D2Ddif}).
The case $\alpha =2$, $\Delta = 0$, and $G_K(x)=KG(x)$  
corresponds to the equation whose integration yields the regular
universal map. 

Integration of Eq.~(\ref{UM1D2Ddif}) with the Riemann-Liouville fractional
derivative $_0D^{\alpha}_tx(t)$ and the initial conditions  
\begin{equation}
(_0D^{\alpha-k}_tx)(0+)=c_k, 
\label{ic}
\end{equation}
where $k=1,...,N$ and $N=\lceil \alpha \rceil$, yields the
Riemann-Liouville universal $\alpha$-FM Eq.~(\ref{EqDir}).

Integration of Eq.~(\ref{UM1D2Ddif}) with the Caputo fractional derivative
$_0^CD^{\alpha}_t x(t)$ and the initial conditions
$(D^{k}_tx)(0+)=b_k$,  
$k=0,...,N-1$ yields the Caputo  universal $\alpha$-FM 
{\setlength\arraycolsep{0.5pt}
\begin{eqnarray}
&&x_{n+1}= \sum^{N-1}_{k=0}\frac{b_k}{k!}h^k(n+1)^{k} 
\nonumber \\  
&&-\frac{h^{\alpha}}{\Gamma(\alpha)}\sum^{n}_{k=0} G_K(x_k) (n-k+1)^{\alpha-1}.
\label{FrCMapx}
\end{eqnarray} 

 In this paper we'll refer to the map
Eqs.~(\ref{EqDir}), the RL universal $\alpha$-FM, as
the Riemann-Liouville universal map with power-law memory or 
the Riemann-Liouville universal fractional map; 
we'll call the Caputo universal $\alpha$-FM, Eq.~(\ref{FrCMapx}),     
the Caputo universal map with power-law memory or 
the Caputo universal fractional map.

In the case of integer $\alpha$ the universal map converges to
$ x_n=0 $ for $\alpha = 0$ and $x_{n+1}=x_n-hG_K(x_n)$ for $\alpha = 1$.
and for $\alpha=N = 2$ with $p_{n+1}=(x_{n+1}-x_n)/h $
\begin{equation}
\begin{array}{c}
\left\{
\begin{array}{lll}
p_{n+1}=p_n-h G_K(x_{n}), \ \ n \ge 0,
\\ 
x_{n+1} = x_{n}+hp_{n+1}, \ \ n \ge 0.
\end{array}
\right.
\end{array} 
\label{IntFr2}
\end{equation} 
N-dimensional, with  $N \ge 2$, universal maps are investigated in \cite{Chaos},
where it is shown that they are volume preserving.

\subsection{Universal fractional difference map}
\label{SecUFDM}

In what follows we will consider fractional Caputo difference maps - 
the only fractional difference maps which behavior has been investigated.  
The following theorem \cite{Chaos2014,ME9,ME10,Fall,DifSum} is essential to 
derive the universal fractional difference map 
%
\begin{theorem}
 For $\alpha \in \mathbb{R}$, $\alpha \ge 0$ the Caputo-like 
h-difference equation 
\begin{equation}
(_0\Delta^{\alpha}_{h,*} x)(t) = -G_K(x(t+(\alpha-1)h)),
\label{LemmaDif_n_h}
\end{equation}
where $t\in (h\mathbb{N})_{m}$, with the initial conditions 
 \begin{equation}
(_0\Delta^{k}_h x)(0) = c_k, \ \ \ k=0, 1, ..., m-1, \ \ \ 
m=\lceil \alpha \rceil
\label{LemmaDifICn_h}
\end{equation}
is equivalent to the map with $h$-factorial-law memory
{\setlength\arraycolsep{0.5pt}   
\begin{eqnarray} 
&&x_{n+1} =   \sum^{m-1}_{k=0}\frac{c_k}{k!}((n+1)h)^{(k)}_h 
\nonumber \\
&&-\frac{h^{\alpha}}{\Gamma(\alpha)}  
\sum^{n+1-m}_{s=0}(n-s-m+\alpha)^{(\alpha-1)} 
G_K(x_{s+m-1}), 
\label{FalFacMap_h}
\end{eqnarray}
}
where $x_k=x(kh)$, 
which is called the $h$-difference Caputo universal  
$\alpha$-family of maps.
\label{T2}
\end{theorem}
In the case of integer $\alpha$ the fractional difference universal map 
converges to 
$x_{n+1}=-G_K (x_n)$ for $\alpha = 0$, $x_{n+1}=x_n-hG_K(x_n)$ for 
$\alpha = 1$,
and for $\alpha=N = 2$ with $p_{n+1}=(x_{n+1}-x_n)/h $
\begin{equation}
\begin{array}{c}
\left\{
\begin{array}{lll}
p_{n+1}=p_n-h G_K(x_{n}), \ \ n \ge 1, \ \ p_1=p_0,
\\ 
x_{n+1} = x_{n}+h p_{n+1},  \ \ n \ge 0.
\end{array}
\right.
\end{array} 
\label{IntFrD2}
\end{equation} 
N-dimensional, with $N \ge 2$, difference universal maps are
volume preserving \cite{Chaos2014}.

All above considered universal maps in the case $\alpha=2$ yield the standard map if $G_K(x)=K\sin(x)$ (harmonic nonlinearity) and we'll call them 
the standard $\alpha$-families of maps.
When $G_K(x)=x-Kx(1-x)$ (quadratic nonlinearity) 
in the one-dimensional case all maps
yield the regular logistic map and we'll call them the logistic 
$\alpha$-families of maps.

\section{Period two sinks and stability of fixed points}
\label{sec:3}

In fractional systems not only the speed of convergence of trajectories to the periodic sinks but also the way in which convergence occurs depends on the initial conditions. As $n \rightarrow \infty$, all trajectories in 
Fig.~2
converge to the same period two ($T=2$) sink (as in Fig.~2~c),
but for small 
values of the initial conditions $x_0$ all
trajectories first converge to the $T=1$ trajectory which then 
bifurcates and turns into the $T=2$ sink converging to its limiting value. 
As $x_0$ increases, the bifurcation point $n_{bif}$ gradually evolves from 
the right to the left (Fig.~\ref{fig:brT2}(a)).   
Ignoring this feature may result (as in 
\cite{Fall} and some other papers) 
in very messy bifurcation diagrams. 

\begin{figure*}[!t]
\begin{center}
\includegraphics[width=0.9\textwidth]{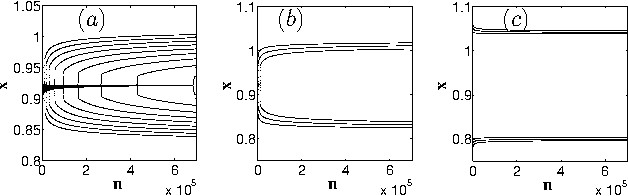}
\vspace{-0.25cm}
\caption{Asymptotically period two trajectories for the Caputo logistic
 $\alpha$-family of maps with  $\alpha=0.1$ and $K=15.5$: (a) nine
 trajectories with the initial conditions $x_0=0.29+0.04i$, $i=0,1,...,8$  
($i=0$ corresponds to the rightmost bifurcation);
(b) $x_0=0.61+0.06i$, $i=1,2,3$; (c) $x_0=0.95+0.04i$, $i=1,2,3$. As $ n
\rightarrow \infty $ all trajectories converge to the limiting values
$x_{lim1}=0.80629$ and  $x_{lim2}=1.036030$ (see Eq~(\ref{eqT2logSolu})). 
The unstable fixed point is $x_{lim0}=(K-1)/K=0.93548$.
}
\end{center}
\label{fig:brT2}
\end{figure*}
In this paper we consider the asymptotic stability of
periodic points. A periodic point is asymptotically stable if there exists
an open set such that all trajectories with initial conditions from this set  
converge to this point as $t \rightarrow \infty$. It is known from the
study of the ordinary nonlinear dynamical systems that as a
nonlinearity parameter increases the system bifurcates. This means that
at the point (value of the parameter) of birth of the $T=2^{n+1}$ sink, the  
$T=2^{n}$ sink becomes unstable. In this section we will investigate the
$T=2$ sinks of discrete fractional systems and apply our results to
analyze stability of the systems' fixed points.

\subsection{$0<\alpha<1$}

When $0<\alpha<1$, all introduced in this paper forms of the universal $\alpha$-family of maps, 
Eqs.~(\ref{EqDir}),
~(\ref{FrCMapx}),~(\ref{FalFacMap_h}),
can be written in the form
\begin{eqnarray}
x_{n+1}= x_0 
-\sum^{n}_{k=0} \tilde{G}(x_k) U(n-k+1).
\label{FrUUMap}
\end{eqnarray} 
In this formula $\tilde{G}(x)=h^\alpha G_K(x)/\Gamma(\alpha)$ and $x_0$ is the initial condition ($x_0=0$ in  Eq.~(\ref{EqDir})).
In fractional maps, Eqs.~(\ref{EqDir})~and~(\ref{FrCMapx}),
\begin{eqnarray}
U_{\alpha}(n)=n^{\alpha-1}, \ \ \ \  U_{\alpha}(1)=1
\label{UnFr}
\end{eqnarray} 
and in fractional difference maps, Eq.~(\ref{FalFacMap_h}),
{\setlength\arraycolsep{0.5pt}
\begin{eqnarray}
&&U_{\alpha}(n)=(n+\alpha-2)^{(\alpha-1)} 
, \ \ \ \  \nonumber \\  
&&U_{\alpha}(1)=(\alpha-1)^{(\alpha-1)}=\Gamma(\alpha).
\label{UnFrDif}
\end{eqnarray} 
}

For $n=2N$ Eq.~(\ref{FrUUMap}) can be written (after subtracting $x_{2N}$) as 
\begin{eqnarray}
&&x_{2N+1} = x_{2N} - \tilde{G}(x_{2N}) U_{\alpha}(1) 
\nonumber \\
&&+\sum^{N}_{n=1} \tilde{G}(x_{2N-2n+1}) \Bigl(U_{\alpha}(2n-1)-U_{\alpha}(2n) \Bigr)
\nonumber \\
&&+\sum^{N}_{n=1} \tilde{G}(x_{2N-2n}) \Bigl( U_{\alpha}(2n)-U_{\alpha}(2n+1) \Bigr).
\label{FrUUMap2Np1}
\end{eqnarray} 
The terms $U_{\alpha}(2n-1)-U_{\alpha}(2n)$ are of the order $n^{\alpha-2}$. If we assume that in the limit $n \rightarrow \infty$ period $T=2$ sink exists, 
\begin{equation}
x_o=\lim_{n \rightarrow
  \infty} x_{2n+1},
\ \ \ \ x_e=\lim_{n \rightarrow
  \infty} x_{2n},
\label{T2pointEx}
\end{equation}
then the series in Eq.(\ref{FrUUMap2Np1}) converge absolutely. In the limit $n \rightarrow \infty$   Eq.(\ref{FrUUMap2Np1}) converges to
\begin{equation}
x_{o} - x_{e} = \Bigl[ \tilde{G}(x_{o})-\tilde{G}(x_{e}) \Bigr] W_{\alpha},
\label{T2_1}
\end{equation} 
where $W_{\alpha}$ is a converging series
\begin{equation}
W_{\alpha}=\sum^{\infty}_{n=1}\Bigl[ U_{\alpha}(2n-1)-U_{\alpha}(2n) \Bigr], 
\label{W}
\end{equation} 
which can be computed numerically with $U_{\alpha}(n)$ defined either by Eq.~(\ref{UnFr}) or by Eq.~(\ref{UnFrDif}). 

Now, instead of subtracting, lets add $x_{2N}$ to $x_{2N+1}$: 
{\setlength\arraycolsep{0.5pt}   
\begin{eqnarray}
&&x_{2N+1} + x_{2N} = 2x_0 
-\sum^{2N}_{n=1} \Bigl[\tilde{G}(x_{2N-n+1})+ 
\nonumber \\ 
&&\tilde{G}(x_{2N-n})\Bigr] U_{\alpha}(n)
- \tilde{G}(x_{0}) U_{\alpha}(2N+1). 
\label{FrUXsum}
\end{eqnarray} 
}
If $T=2$ sink exists, then, in the limit $n \rightarrow \infty$, the
left-hand side (LHS) of Eq.~(\ref{FrUXsum}), as well as the first term on the right-hand side (RHS) and the last term of this equation, is finite.
Expressions in the brackets in Eq.~(\ref{FrUXsum}) tend to the limit $\tilde{G}(x_o)+\tilde{G}(x_e)$. Because the series $\sum^{\infty}_{n=1}U_{\alpha}(n)$ is diverging, the only case in which Eq.~(\ref{FrUXsum}) can be true is when
\begin{equation}
\tilde{G}(x_o)+\tilde{G}(x_e)=0.
\label{Req2}
\end{equation}
Equations which define the existence and value of the asymptotic $T=2$ 
sink can be written as   
\begin{equation}
\begin{array}{c}
\left\{
\begin{array}{lll}
G_K(x_o)+G_K(x_e)=0,
\\ 
x_{o} - x_{e} = \frac{W_{\alpha}}{\Gamma(\alpha)} h^{\alpha}\Bigl[ G_K(x_{o})-G_K(x_{e}) \Bigr] .
\end{array}
\right.
\end{array} 
\label{UMapT2sink}
\end{equation} 

\begin{itemize}
\item
{
It is easy to see that the fixed point, defined by the equation
$G_K(x_{o})=0$ is a solution of the system Eq.~(\ref{UMapT2sink}).
}
\item
{
As it was mentioned above, when $h \rightarrow 0$, fractional difference 
equations converge to the corresponding fractional differential
equations. As $h \rightarrow 0$, the second equation from the system 
Eq.~(\ref{UMapT2sink}) leads to $x_{o}-x_{e} \rightarrow 0$. 
This implies that in fractional differential equations of the order 
$0<\alpha<1$ transition from a fixed point to periodic trajectories 
will never happen. A strict proof of the impossibility of periodic 
trajectories (except fixed points) in autonomous fractional systems 
described by the fractional differential equation 
\begin{equation}
\frac{d^{\alpha}x}{dt^{\alpha}}=G_K(x(t)), \  \  \  0<\alpha<1 
\label{FDE}
\end{equation}
with the Caputo or Riemann-Liouville fractional derivative was given in 
\cite{PerC1} (Theorem 9 there). Nonexistence of periodic trajectories and 
the fact that in regular dynamics transition to chaos occurs through 
cascades of the period doubling bifurcations, leads us to the following 
conjecture
\begin{conj}
Chaos does not exist in continuous fractional systems of the 
orders $0<\alpha<1$. 
\label{C1}
\end{conj}
}
\end{itemize}

\subsection{$1<\alpha<2$}

For $1<\alpha<2$ map equations 
Eqs.~(\ref{EqDir}),~(\ref{FrCMapx}),~(\ref{FalFacMap_h}),
can be written in the form 
{\setlength\arraycolsep{0.5pt}
\begin{eqnarray}
&&x_{n+1}= x_0 +f(\alpha)[h(n+1)]^{\beta}p_0  
\nonumber \\  
&&-h\sum^{n}_{k=0} \tilde{G}(x_k) U_{\alpha}(n-k+1)+hf_1(n).
\label{FrUUMap2x}
\end{eqnarray} 
}
Here $\tilde{G}(x)=h^{\alpha-1} G_K(x)/\Gamma(\alpha)$, $x_0$ and $U(n)$ 
are defined the same way as in   
Eqs.~(\ref{FrUUMap}),~(\ref{UnFr}),~and~(\ref{UnFrDif}), $p_0$ is the initial 
momentum ($b_k$ or $c_k$ in corresponding formulae), $\beta$ is equael to
$1$ in  
Eqs.~(\ref{FrCMapx}),~(\ref{FalFacMap_h}) and $\alpha-1$ in 
Eq.~(\ref{EqDir}) $f(\alpha)$ is $1$ in 
Eqs.~(\ref{FrCMapx}),~(\ref{FalFacMap_h}) and $1/\Gamma(\alpha)$ in 
Eq.~(\ref{EqDir}), and $f_1(n)=0$ 
in Eqs.~(\ref{EqDir}),~(\ref{FrCMapx}) and 
$f_1(n)=h^{\alpha-1} G(x_0)(n-1+\alpha)^{(\alpha-1)}/\Gamma(\alpha) \sim n^{\alpha-1}$ in 
Eq.~(\ref{FalFacMap_h}).

If we define 
\begin{eqnarray}
p_{n+1}= \frac{x_{n+1}-x_{n}}{h}, 
\label{p}
\end{eqnarray} 
then, taking into account that $U_{\alpha}(0)=0$, from Eq.~(\ref{FrUUMap2x}) follows
{\setlength\arraycolsep{0.5pt}
\begin{eqnarray}
&&p_{n+1}=\tilde{f}(n) p_0 
\nonumber \\  
&&-\sum^{n}_{k=0}   \tilde{G}(x_k) \tilde{U}_{\alpha}(n-k+1)
+f_1(n)-f_1(n-1),
\label{FrUUMap2p}
\end{eqnarray}
} 
where
{\setlength\arraycolsep{0.5pt}   
\begin{eqnarray}
\tilde{U}_{\alpha}(n)=U_{\alpha}(n)-U_{\alpha}(n-1 )\nonumber
\\ 
=
\begin{array}{c}
\left\{
\begin{array}{lll}
n^{\alpha-1}-(n-1)^{\alpha-1} \sim n^{\alpha-2}
\\ {\rm and} \ \ \tilde{U}_{\alpha}(1)=1 
\  \    
{\rm in \ \ Eqs.~(\ref{EqDir}),~(\ref{FrCMapx})} ;
\\ 
(n+\alpha -2)^{(\alpha -1)}-(n+\alpha -3)^{(\alpha -1)} 
\\=(\alpha-1)(n+\alpha -3)^{(\alpha-2)} \\
=(\alpha-1)U_{\alpha-1}(n)
\sim n^{\alpha-2} 
\\  {\rm and} \ \ 
\tilde{U}_{\alpha}(1)=\Gamma(\alpha)
\  \  {\rm in \ \
  Eq.~(\ref{FalFacMap_h})},
\end{array}
\right.
\end{array} 
\label{Utilde}
\end{eqnarray} 
}
$f_1(n)-f_1(n-1)=0$  
in Eqs.~(\ref{EqDir}),~(\ref{FrCMapx}) and 
$f_1(n)- f_1(n-1) \sim n^{\alpha-1}$ in 
Eq.~(\ref{FalFacMap_h}), $\tilde{f} (n)=1$ in
Eqs.~(\ref{FrCMapx}),~(\ref{FalFacMap_h}) and  
$\tilde{f}(n) \sim n^{\alpha-2}$ in  Eq.~(\ref{EqDir}).
Note, that the definition of $\tilde{U}_{\alpha}(1)$ in Eq.~(\ref{Utilde}) and 
$U_{\alpha}(1)$ in Eqs.~(\ref{UnFr}),~(\ref{UnFrDif}) are identical. 

Assuming existence of the $T=2$ sink and limits  $x_o$ and $x_e$ are  
defined by Eq.~(\ref{T2pointEx}), the limiting values for $p$ are defined
by
{\setlength\arraycolsep{0.5pt}   
\begin{eqnarray}
&&p_o=\lim_{n \rightarrow
  \infty} p_{2n+1}= \lim_{n \rightarrow
  \infty} \frac{x_{2n+1}-x_{2n}}{h}=\frac{x_o-x_e}{h} \nonumber \\ 
&&{\rm and} \ \ p_e=\lim_{n \rightarrow
  \infty} p_{2n}=-p_o.
\label{T2pointEp}
\end{eqnarray} 
}
As in the derivation of Eqs.~(\ref{T2_1})~and~(\ref{Req2}), if we add and
subtract expressions for $p_{2N+1}$ and $p_{2N}$, we'll arrive at relations
\begin{equation}
p_{o} - p_{e} = \Bigl[ \tilde{G}(x_{o})-\tilde{G}(x_{e}) \Bigr]
\tilde{W}_{\alpha} 
\label{T2_1p}
\end{equation}
and 
\begin{equation}
\tilde{G}(x_o)+\tilde{G}(x_e)=0,
\label{Req2p}
\end{equation} 
where $\tilde{W}_{\alpha}$ is a converging series
\begin{equation}
\tilde{W}_{\alpha}=\sum^{\infty}_{n=1}\Bigl[ \tilde{U}_{\alpha}(2n-1)- \tilde{U}_{\alpha}(2n) \Bigr]. 
\label{tW}
\end{equation}
Let's note that with $U_{\alpha}(n)=n^{\alpha-1}$, as defined in 
Eq.~(\ref{UnFr}), 
$\tilde{W}$ is identical to the introduced in \cite{ME2}
$V_{\alpha l}$ defined as
\begin{equation}
\tilde{W}_{\alpha}=V_{\alpha l}=\sum^{\infty}_{n=1}(-1)^{n+1}
\Bigl[ n^{\alpha-1}- (n-1)^{\alpha-1} \Bigr]. 
\label{tV}
\end{equation}
High accuracy algorithm for calculating 
$V_{\alpha l}$ is presented in Appendix to \cite{ME4}.   For 
$U(n)$ defined by Eq.~(\ref{UnFrDif}) $\tilde{W}$ was calculated in 
\cite{Chaos2014}. Taking into account that converging series Eq.~(\ref{tW})
can be written as 
\begin{equation}
\tilde{W}_{\alpha}= \tilde{U}_{\alpha 1} - \sum^{\infty}_{n=1}\Bigl[ \tilde{U}_{\alpha}(2n)- \tilde{U}_{\alpha}(2n+1) \Bigr], 
\label{tWn}
\end{equation}
where 
{\setlength\arraycolsep{0.5pt}   
\begin{eqnarray}
\tilde{U}_{\alpha 1}
=
\begin{array}{c}
\left\{
\begin{array}{lll}
1 \ \    
{\rm in \ \ Eqs.~(\ref{EqDir}),~(\ref{FrCMapx})} ,
\\ 
\Gamma(\alpha)
\ \  {\rm in \ \
  Eq.~(\ref{FalFacMap_h})},
\end{array}
\right.
\end{array} 
\label{U1}
\end{eqnarray} 
}
and using absolute convergence of series Eq.~(\ref{W}) (and,
correspondingly, the series on the first line of Eq.~(\ref{tWall})
below), for $0<\alpha <1$
we can write
\begin{eqnarray}
&&\tilde{W}_{\alpha}= \tilde{U}_{\alpha 1} 
-\sum^{\infty}_{n=1} \Bigl\{  \Bigl[
{U}_{\alpha}(2n)- {U}_{\alpha}(2n-1) \Bigr]-  
\nonumber \\ 
&&-\Bigl[ {U}_{\alpha}(2n+1)- {U}_{\alpha}(2n) \Bigr] \Bigr\}
\nonumber \\ 
&&=\tilde{U}_{\alpha 1} + \sum^{\infty}_{n=1} \Bigl[
{U}_{\alpha}(2n-1)- {U}_{\alpha}(2n) \Bigr]
\\ 
&&- \sum^{\infty}_{n=1} \Bigl[ {U}_{\alpha}(2n)- {U}_{\alpha}(2n+1)
\Bigr] =W_{\alpha} + \tilde{U}_{\alpha 1} 
\nonumber \\ 
&&-U_{\alpha}(2)+U_{\alpha}(3)-U_{\alpha}(4)+U_{\alpha}(5)-...=2W_{\alpha}. \nonumber
\label{tWall}
\end{eqnarray} 
Let us notice that in fractional difference maps Eq.~(\ref{tWn}) can be written as
{\setlength\arraycolsep{0.5pt}   
\begin{eqnarray}
&&\tilde{W}_{\alpha}= (\alpha-1)\Gamma(\alpha-1) - (\alpha-1) \sum^{\infty}_{n=1}\Bigl[ U_{\alpha-1}(2n)- 
\nonumber \\ 
&&U_{\alpha-1}(2n+1) \Bigr]
=(\alpha-1)W_{\alpha-1}=\frac{\alpha-1}{2} \tilde{W}_{\alpha-1}. 
\label{tWnFD}
\end{eqnarray}
}

Finally, the equations which define the existence and value of 
the asymptotic $T=2$ sink for $0 < \alpha <2$ can be written as   
\begin{equation}
\begin{array}{c}
\left\{
\begin{array}{lll}
G_K(x_o)+G_K(x_e)=0,
\\ 
x_{o} - x_{e} = \frac{\tilde{W}_{\alpha}}{2\Gamma(\alpha)} h^{\alpha}\Bigl[ G_K(x_{o})-G_K(x_{e}) \Bigr] ,
\end{array}
\right.
\end{array} 
\label{UMapT2sink2}
\end{equation} 
where $\tilde{W}_{\alpha}$ is defined by
Eqs.~(\ref{tWn}),~(\ref{U1}). Notice that according to Eq~(\ref{tWn}) $\tilde{W}_1=1$.
 \begin{itemize}
\item
{
As in the case  $0 < \alpha <1$, for   $1 < \alpha <2$ the fixed point, 
defined by the equation
$G_K(x_{o})=0$ is a solution of the system Eq.~(\ref{UMapT2sink2}).
}
\item
{
As $h \rightarrow 0$, fractional difference 
equations converge to the corresponding fractional differential
equations and $x_{o}-x_{e} \rightarrow 0$, 
which implies that in fractional differential equations of the order 
$1<\alpha<2$ transition from a fixed point to periodic trajectories 
will never happen.  Now we may formulate a stronger conjecture:
\begin{conj}
Chaos does not exist in continuous fractional systems of the 
orders $0<\alpha<2$. 
\label{C2}
\end{conj}
} 
\end{itemize}

\subsection{Examples}

Now we'll consider application of the results from
this section to the
introduced at the end of Section~\ref{sec:2} fractional
and fractional difference standard ($G_K(x)=K \sin(x)$) and
logistic ($G_K(x)=x-Kx(1-x)$ ) $\alpha$-families of maps.

\subsubsection{Standard $\alpha$-families of maps}

With $G_K(x)=K \sin(x)$ all above considered forms of the universal map
for $\alpha=2$ converge to the regular standard map and they are called
the standard $\alpha$-families of maps. These families of maps are usually
considered on a torus (mod $2\pi$). The first equation of the system 
Eq.~(\ref{UMapT2sink2}) yields 
\begin{equation}
\sin \frac{x_o+x_e}{2} \cos \frac{x_o-x_e}{2} = 0, 
\label{stCond1}
\end{equation}
which on $x \in [-\pi, \pi]$ yields two solutions
{\setlength\arraycolsep{0.5pt}
\begin{eqnarray}
&&{\rm symmetric \ \ point} \  \ x_{osy}=-x_{esy} \  \ {\rm and} \  \nonumber  
\\
&&{\rm shift}-\pi  \ \ {\rm point} \  \ x_{osh}=x_{esh}-\pi. 
\label{stCond1solve}
\end{eqnarray}
}
Then, the second equation of  Eq.~(\ref{UMapT2sink2}) yields 
the equation which together with Eq.~(\ref{stCond1solve}) 
defines two $T=2$ sinks for $0<\alpha<2$
\begin{equation}
\sin x_{osy}= \frac{2 \Gamma(\alpha)}{\tilde{W}_{\alpha}
  h^{\alpha}K}x_{osy} 
\label{stCond2b}
\end{equation}
and
\begin{equation}
\sin x_{osh}= \frac{\pi \Gamma(\alpha)}{\tilde{W}_{\alpha} h^{\alpha}K}.
\label{stCond2d}
\end{equation}

The symmetric $T=2$ sink appears when  
\begin{equation}
h^{\alpha}|K|>h^{\alpha}|K_{C1s}|=\frac{2\Gamma(\alpha)}{\tilde{W}_{\alpha}}
\label{ExistStCond1b}
\end{equation}
and the shift-$\pi$ $T=2$ sink appears when 
\begin{equation}
h^{\alpha}|K|>\frac{\pi}{2}h^{\alpha}|K_{C1s}|.
\label{ExistStCond1c}
\end{equation}
\begin{figure*}[!t]
\begin{center}
\includegraphics[width=0.8\textwidth]{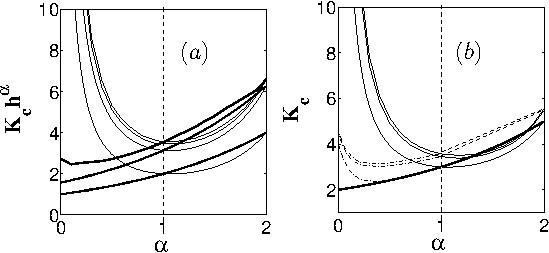}
\vspace{-0.25cm}
\caption{2D bifurcation diagrams for fractional (solid thing lines) and
fractional difference (bold and dashed lines) Caputo standard (a) 
and $h=1$ logistic (b) maps. Fist bifurcation, transition from the stable fixed 
point to the stable period two ($T=2$) sink, 
occurs on the bottom curves.
$T=2$ sink (in the case of standard $\alpha$-families of maps
antisymmetric T=2 sink with $x_{n+1}=-x_n$)
is stable between the bottom and the middle curves. Transition 
to chaos occurs on the top curves. For the standard fractional map transition from $T=2$ to $T=4$ sink
occurs on the line below the top line (the third from the bottom line). 
Period doubling bifurcations leading to
chaos occur in the narrow band between the two top curves.
All bottom curves, as well as the next to the bottom in (a), are obtained using formulae 
Eqs.~(\ref{ExistStCond1b}),~(\ref{ExistStCond1c}),~and~(\ref{condT2log}).
Two dashed lines for $1<\alpha <2$ in (b) are obtained by
interpolation. The remaining lines are results of the direct
numerical simulations. Stability of the fixed point for the fractional difference logistic 
$\alpha$-family of maps is calculated using both, Eq.~(\ref{condT2log}) (bold solid line) and
the direct numerical simulations (a dashed line branching from the solid line). The
difference is due to the slow, as $n^{-\alpha}$ (see \cite{Chaos2014}), convergence of trajectories to 
the $T=2$ sink for small $\alpha$ ($x$ vs. $K$, fixed $\alpha$, bifurcation diagrams used to find 
the first bifurcation were calculated on trajectories after 5000 iterations).   
}
\end{center}
\label{BD2d}
\end{figure*}

\subsubsection{Logistic $\alpha$-families of maps}

With $G_K(x)=x-K x(1-x)$ all above considered forms of the universal map
for $\alpha=1$ converge to the regular logistic map and they are called
the logistic $\alpha$-families of maps. Th system Eq.~(\ref{UMapT2sink2})
becomes
\begin{equation}
\begin{array}{c}
\left\{
\begin{array}{lll}
(1-K)(x_o+x_e)+K(x_o^2+x_e^2)=0,
\\ 
x_{o} - x_{e} = \frac{\tilde{W}_{\alpha}}{2\Gamma(\alpha)}
h^{\alpha}  (x_o-x_e)[1-K+(x_o+x_e)] 
\end{array}
\right.
\end{array} 
\label{T2Logi}
\end{equation} 
Two fixed point solutions with $x_o=x_e$ are $x_o=0$, stable for
$K<1$, and $x_o=(K-1)/K$.

The $T=2$ sink is defined by the equation 
\begin{equation}
x_o^2-\Bigl( \frac{2\Gamma(\alpha)}{\tilde{W}Kh^{\alpha}} 
+\frac{K-1}{K}  \Bigr)x_o +
\frac{2\Gamma^2(\alpha)}{(\tilde{W}Kh^{\alpha})^2}
+\frac{(K-1)\Gamma(\alpha)}{\tilde{W}K^2h^{\alpha}}=0,
\label{eqT2log}
\end{equation}
which has solutions 
\begin{equation}
x_o=\frac{K_{C1s}+K-1 \pm \sqrt{(K-1)^2-K_{C1s}^2}}{2K}
\label{eqT2logSolu}
\end{equation}
defined when
\begin{equation}
K \ge 1 + \frac{2\Gamma(\alpha)}{\tilde{W}h^{\alpha}}=1+K_{C1s} \ \ {\rm or} \ \
K \le 1 - \frac{2\Gamma(\alpha)}{\tilde{W}h^{\alpha}}=1-K_{C1s} 
\label{condT2log}
\end{equation}
The first inequality of Eq.~(\ref{condT2log}) was derived in \cite{ME4} for
$h=1$ and $1 < \alpha <2 $.
In this paper we consider $K>0$ and $h \le 1$. It follows from the definition,
Eq.~(\ref{tWn}), that  $\tilde{W}<\tilde{U}_1$, which is either 1 or
$\Gamma(\alpha)$, and it is known that $\Gamma(\alpha)>0.885$ for $\alpha>0$.
Then, $ 2\Gamma(\alpha)/(\tilde{W}h^{\alpha})>1$ and  we may ignore the
second  of the inaquolities in Eq.~(\ref{condT2log}). We may also note that
the fixed point $x=(K-1)/K$ is stable when
\begin{equation}
1 \le K <K_{C1l}= 1 + \frac{2\Gamma(\alpha)}{\tilde{W}h^{\alpha}}=1+K_{C1s}.
\label{condT2logn}
\end{equation}

\section{Conclusion}
\label{sec:4}

Figs.~3~a~and~b, the two-dimensional bifurcation diagrams, 
present results of the computer simulations
of the fractional and fractional difference standard and logistic maps.
Low curves on these diagrams are obtained using  
Eqs.~(\ref{ExistStCond1b}),~(\ref{ExistStCond1c}),~and~(\ref{condT2log}). 
They are in good agreement with the results (also used to calculate all
other curves) obtained by the direct numerical simulations 
by calculating $x$ vs. $K$ bifurcation 
diagrams for various $\alpha \in (0,2) $ after 5000 iterations.
Slight difference in Fig.~3~b for the fractional difference
logistic map for $\alpha<0.2$ is probably due to the slow, 
as $\sim n^{-\alpha}$ convergence of trajectories to the fixed points. 
This confirms the validity of Eq.~(\ref{UMapT2sink2}) to calculate 
the coordinates of the asymptotic $T=2$ sinks and the points of the first  
bifurcations for the discrete fractional/fractional difference maps.
The continuous limits of the considered in this paper discrete maps
are fractional differential equations and from the consideration presented in
this paper we may conclude that chaos is impossible in systems 
described by equations
\begin{equation}
\frac{d^{\alpha}x}{dt^{\alpha}}=f(x)  
\label{Last}
\end{equation}   
with $0<\alpha<2$. 

There are still many unanswered questions related to the behavior of    
fractional systems. They include:
\begin{itemize}
\item
{
What is the nature and the corresponding analytic description of the
bifurcations on a single trajectory of a fractional system?   
}
\item
{
What kind of self-similarity can be found in CBTT?   
}
\item
{
How to describe a self-similar behavior corresponding to the bifurcation 
diagrams of fractional systems? Can constants, similar to the Feigenbaum 
constants be found?   
}
\item
{
Can cascade of bifurcations type trajectories be found in continuous systems?   
}
\end{itemize}

This paper is a small step in investigation of the fractional dynamical
systems and we hope that the following works will lead to more complete 
description of fractional (with power-law memory) systems which have
many applications in biological, social, and physical systems.


\section*{Acknowledgments}
The author acknowledges continuing support from Yeshiva University 
and expresses his gratitude to the administration of Courant
Institute of Mathematical Sciences at NYU
for the opportunity to complete this work at Courant. 


\end{document}